\documentstyle[11pt,aaspp4]{article}

\begin{document}
\newcommand{\decsec}[2]{$#1\mbox{$''\mskip-7.6mu.\,$}#2$}
\newcommand{\decsectim}[2]{$#1\mbox{$^{\rm s}\mskip-6.3mu.\,$}#2$}
\newcommand{\decmin}[2]{$#1\mbox{$'\mskip-5.6mu.$}#2$}
\newcommand{\asecbyasec}[2]{#1$''\times$#2$''$}
\newcommand{\etal}{et al.}

\title{The Optical Afterglow of GRB\,971214: R and J Photometry}

\author{A. Diercks\altaffilmark{1}, E. W. Deutsch\altaffilmark{1}, F.
  J. Castander\altaffilmark{2}, C. Corson\altaffilmark{3}, G.
  Gilmore\altaffilmark{4}, D. Q. Lamb\altaffilmark{2},\\ N.
  Tanvir\altaffilmark{4}, E. L. Turner\altaffilmark{5}, and R.
  Wyse\altaffilmark{6}}

\altaffiltext{1}{Department of Astronomy, Box 351580, University of
  Washington, Seattle, WA 98195; diercks@astro.washington.edu;
  deutsch@astro.washington.edu}

\altaffiltext{2}{Department of Astronomy and Astrophysics, University
  of Chicago, 5640 South Ellis Avenue, Chicago, IL 60637;
  fjc@oddjob.uchicago.edu; lamb@pion.uchicago.edu}

\altaffiltext{3}{Apache Point Observatory, 2001 Apache Point Rd.,
  Sunspot NM 88349; corson@galileo.apo.nmsu.edu}

\altaffiltext{4}{Institute of Astronomy, Madingley Road, Cambridge CB3
  0HA, UK; gil@ast.cam.ac.uk}

\altaffiltext{5}{Princeton University Observatory, Peyton Hall,
  Princeton, NJ 08544; elt@astro.princeton.edu}

\altaffiltext{6}{Department of Physics and Astronomy, The Johns
  Hopkins University, Baltimore, MD 21218; wyse@wyser.pha.jhu.edu}

\begin{center}
Accepted for publication in The Astrophysical Journal Letters\\
Volume 503, L105\\
{\it received 1998 March 23; accepted 1998 June 24}
\end{center}

\begin{abstract}
  We present an $R$-band and $J$-band photometry of an optical
  transient which is likely to be associated with the gamma-ray burst
  event GRB\,971214.  Our first measurement took place 13 hours after
  the gamma-ray event.  The brightness decayed with a power-law
  exponent $\alpha = -1.20 \pm 0.02$, which is similar to those of
  GRB\,970228 and GRB\,970508 which had exponents of $\alpha = -1.10
  \pm 0.04$ and $\alpha = -1.141\pm 0.014$ respectively.  The
  transient decayed monotonically during the first four days following
  the gamma-ray event in contrast with the optical transient
  associated with GRB\,970508 which increased in brightness, peaking
  two days after the burst, before settling to a power-law decay.
\end{abstract}

\clearpage

\section{Introduction}
The launch of the BeppoSAX satellite \cite{boel97} in 1996 has led to
a recent breakthrough in the study of gamma-ray bursts by detecting
fading X-ray counterparts and localizing them to a few arc-minutes on
the sky.  This has allowed subsequent identification of optical
counterparts in three cases, GRB\,970228 \cite{groo97}, GRB\,970508
\cite{bond97}, and GRB\,971214 \cite{halp97}.

GRB\,971214 triggered the BeppoSAX Gamma Ray Burst Monitor on December
14.97272 UT with a peak flux of 650 counts s$^{-1}$.  In addition, the
burst was localized to an error radius of \decmin{3}{9} (99\%
confidence) with the Wide Field Camera \cite{heis97}. The gamma-ray
event was also detected by BATSE which measured a total fluence above
20 keV of $1.09 \pm 0.07 \times 10^{-5}$ erg cm$^{-2}$ and by one
RXTE-ASM camera yielding a peak intensity of $470 \pm 140$ mCrab
\cite{kipp97}. Shortly thereafter, a fading optical source was
detected within the BeppoSAX error circle at
$\alpha(J2000)={\rm11^h56^m}$\decsectim{26}{4},
$\delta(J2000)={\rm65^\circ12'}$\decsec{00}{5} with I-band magnitudes
of $21.2 \pm 0.3$ on Dec 15.47 UT and $\sim 22.6$ on Dec 16.47 UT
\cite{halp97}. Further observations by BeppoSAX detected a previously
unknown fading X-ray source, later designated 1SAX J1156.4+6513,
within the initial error circle at $\alpha(J2000)={\rm11^h56^m}25^{\rm
s}$, $\delta(J2000)={\rm65^\circ13'}11''$ with an error radius of
about $1'$ \cite{anto97}. Since this second X-ray detection is
consistent with the position of the fading optical source identified
by Halpern \etal\ (1997) it is quite likely that these objects are the
X-ray and optical afterglow of GRB\,971214.

We report here $R$-band and $J$-band observations of this optical
transient (OT) made with the Apache Point Observatory(APO) 3.5 m
telescope.  This work extends the preliminary photometry reported in
Diercks \etal\ (1997), Castander \etal\ (1997), and Tanvir \etal\ 
(1997).

\section{$R$-band Observations}
The field of GRB\,971214 was imaged on the four consecutive nights
following the burst, with the first observations taking place 13 hours
after the BeppoSAX detection.  The optical transient was positively
detected each night.  All observations were taken with a thinned SITe
2048 CCD through an $R$-band filter at $2\times2$ binning yielding a
plate scale of \decsec{0}{28} pixel$^{-1}$.  The images were overscan
subtracted and flat-fielded with twilight flats in the usual manner
using IRAF \cite{tody93}.  The field of GRB\,971214 is plagued by the
presence of a bright star HD103690 ($V = 6.7$) which contributed a
substantial amount of stray light to the images.  In order to
optimally combine the images, we removed this stray light by median
filtering each frame with a 14$''$ rectangular box and then smoothing
the resulting image with a 18$''$ FWHM Gaussian filter.  The spatially
varying part of each smoothed image was then subtracted from its
corresponding flat-fielded image to remove the excess stray light.  As
the variations were typically less than 10\%, the effect on the noise
computations due to slightly lower sky values in certain regions of
the images was quite small.

\begin{figure}
\plotone{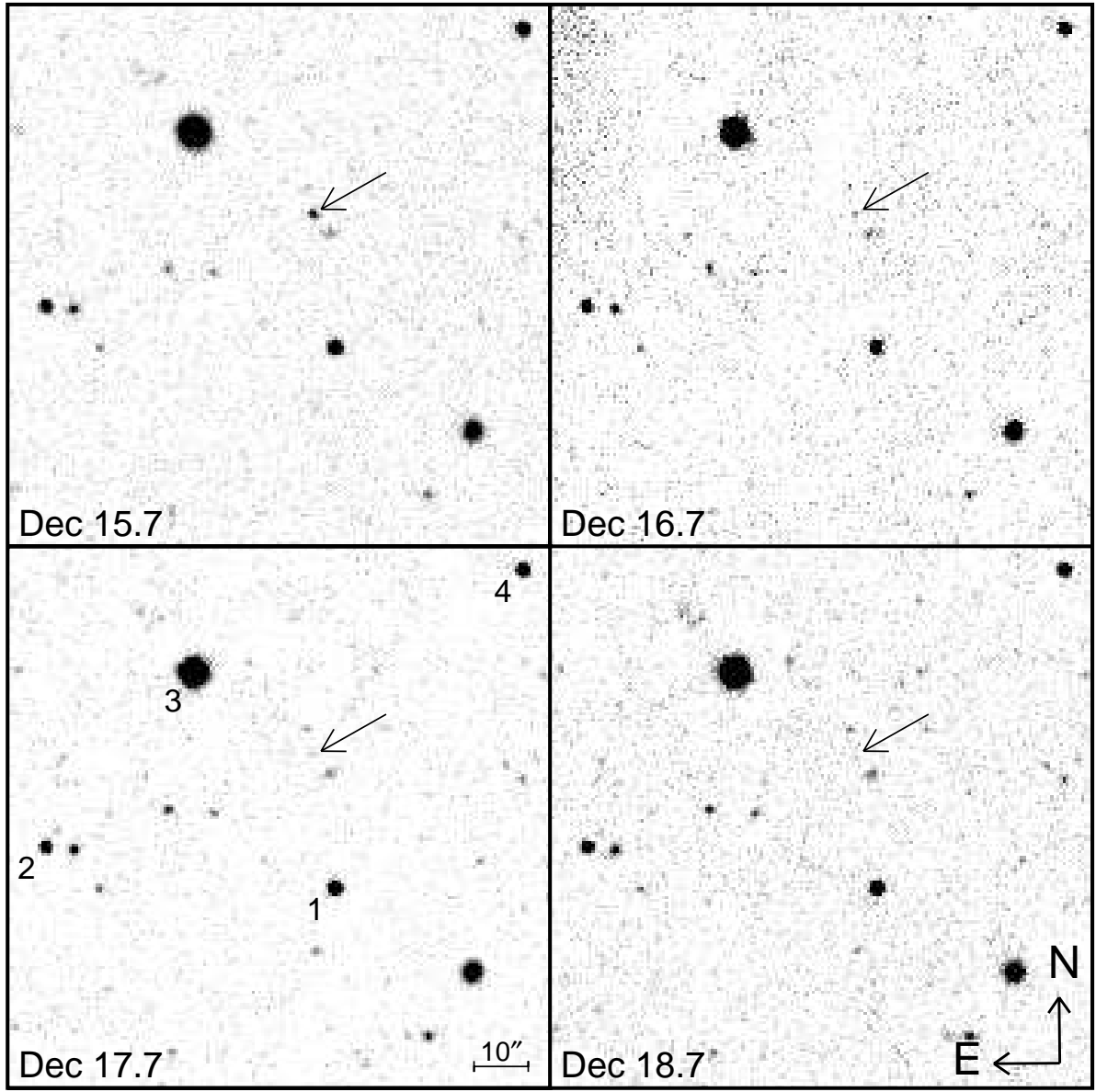}
\caption{\asecbyasec{100}{100} R-band images of the field of
GRB\,971214 on the four nights observed from APO.  The clearly fading
optical transient is indicated with an arrow, and several of the
comparison stars are numbered.\label{fig-1}}
\end{figure}

All the images for each night were aligned and combined, weighting
each frame by the photon noise within a seeing disk.  The epoch quoted
for each measurement is average of the time of each observation
weighted in the same manner.  The four final images are shown in
Figure~\ref{fig-1}.

The OT discovered by Halpern is clearly visible in our December 15.51
data and has sufficient signal to fit an accurate centroid using
DoPhot \cite{sche93}.  We measure a position of
$\alpha(J2000)={\rm11^h56^m}$\decsectim{26}{6},
$\delta(J2000)={\rm65^\circ12'}$\decsec{00}{9} for the OT with an
uncertainty of \decsec{0}{5} using five nearby stars in the USNO-A1.0
catalog \cite{mone96}.  All subsequent photometry on both the OT and
comparison stars was done using DoPhot with these positions fixed as
it resulted in greater precision especially in the later data where
the OT is quite faint.

In order to test the robustness of our detection in the December 18.50
UT data where the OT is close to the detection limit, we performed
fixed position PSF photometry on 50 arbitrary regions within the image.
In only one case was there sufficient signal to fit a PSF, and the
returned magnitude was at the detection limit of the image.  We are
therefore confident in our detection of the OT in the December 18.50
data.

We observed the field PG0918+029 \cite{land93} and the field of
GRB\,971214 on 1998 February 1 under photometric conditions with the
same camera and filter arrangement used to monitor the OT.\ 
Observations were made over a range of airmass from 1.18 to 2.32 and
we derived an average R-band extinction of $0.088 \pm 0.031$ mag
airmass$^{-1}$.

\begin{figure}
\plotone{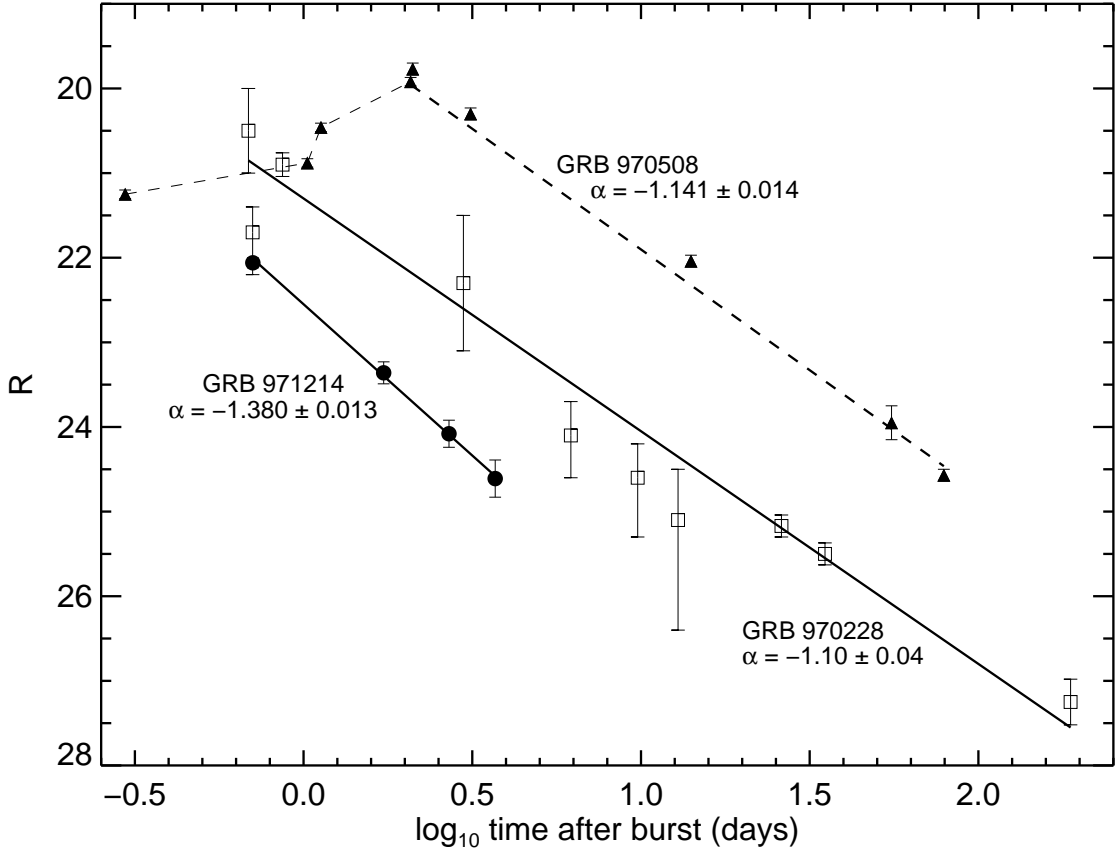}
\caption{$R$-band lightcurve of the optical transient associated with
  GRB\,971214 based on our observations at APO.  The $R$-band
  lightcurve data points and power-law decay fit lines of the
  afterglows of GRB\,970228 ({\it squares})\cite{gala97} and
  GRB\,970508 ({\it triangles}) \cite{gala98} are also shown for
  comparison.  A thin dashed line is drawn between the data points
  measured during the rising phase of the GRB\,970508 lightcurve.
  \label{fig-2}}
\end{figure}

Using this coefficient we obtained calibrated magnitudes for five
stars in the field of GRB971214 which are listed in Table~\ref{tbl-1}
and shown in Figure~\ref{fig-1}.  Star 3 was saturated in some of the
monitoring data and was not used for calibration purposes.  Using the
remaining stars, we normalized the monitoring data to obtain the
absolute photometry of the OT on each night.  The lightcurve for the
four days following the burst is shown in Figure~\ref{fig-2} along
with the lightcurves of the OTs associated with GRB\,970228
\cite{gala97} and GRB\,970508 \cite{gala98} for comparison.

\placetable{tbl-1}

\section{$J$-band Observations}
We also imaged the GRB971214 field with the Grism Spectrometer and
Imager II (GRIM II) in the J band on the first two nights after the
burst.  The GRIM II instrument has a NICMOS array, which at f/5
provides a pixel scale of 0.47$''$ and a field of view of $2' \times
2'$.  We obtained dithered 10-second exposures, totaling 730 and 1150
seconds on the first and second nights, respectively.  Images were
dark-current and bad-pixel corrected, flat-fielded and combined, with
high sigma threshold clipping, into a single image for each night. The
mean time of the resulting exposures were December 15.44 and 16.45 UT,
respectively.

We detect the OT at a magnitude of $J=20.47^{+0.21}_{-0.19}$ on
December 15.44 and $J=21.46^{+0.34}_{-0.26}$ on December 16.45, where the
calibration is performed using observations of the UKIRT faint
standard star FS12, carried out at several airmasses.
Table~\ref{tbl-2} summarizes both the $R$-band and $J$-band
observations.

\placetable{tbl-2}

\section{Discussion}
The well-observed light curve of the GRB\,970508 OT, the brightest
observed thus far, shows a dramatic rise, peaking nearly two days
after the initial burst, before beginning a power-law decay.
GRB\,970228 was not observed nearly as often through a consistent
filter, but there is also evidence (depending on spectral assumptions)
\cite{guar97} that the transient increased in brightness until
$\sim20$ hr after the burst, after which it also began fading with a
power-law slope essentially identical to that of GRB\,970508.  Despite
the same decay slope, the GRB\,970228 OT was $\sim1.5$ mag fainter
than the GRB\,970508 OT.  A detailed analysis of the difference
between these two light curves is presented in Pedersen et al. (1998).

A power-law of the form $F = F_{0}t^{\alpha}$ was fit to the four
$R$-band data points yielding $\alpha = -1.20 \pm 0.02$.  This rate of
decline is similar to the two previously identified bursts although
there is no evidence of the OT brightening over the course of the
observations.

\placefigure{fig-2}

\begin{figure}
\plotone{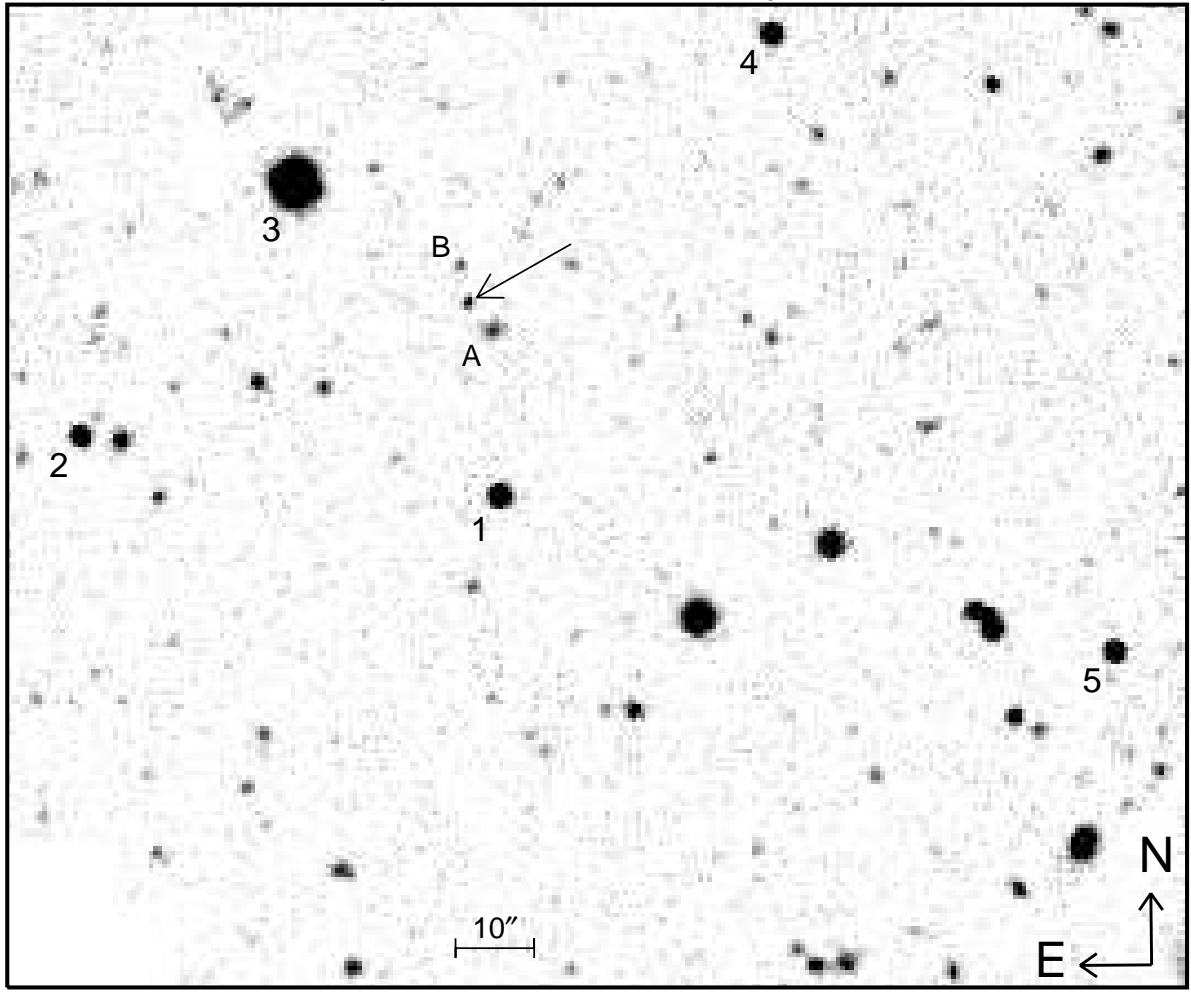}
\caption{\asecbyasec{150}{120} R-band combined deep image of the GRB
  971214 field.  The optical transient is marked with an arrow.
  Photometric reference stars $1-5$ and selected other objects discussed
  in the text are labeled.\label{fig-3}}
\end{figure}

The observations from all four nights are combined into one deep image
totaling 3.25 hrs of integration with a limiting $R$-band magnitude
$\sim25.4$ (Figure~\ref{fig-3}).  The two brightest objects within
20$''$ of the OT are an extended source (A) \decsec{4}{6} southwest of
the transient with $R = 22.7 \pm 0.1$, and an unresolved source (B)
\decsec{4}{9} north of the transient with $R = 23.43 \pm 0.08$.  The
resolution of these images is insufficient to identify any structure
in the extended object. \footnote{To obtain the images discussed in
this work, contact A. Diercks or E. Deutsch}

\section{Acknowledgments}
We would like to thank Jan van Paradjis and the BeppoSAX team for
their critical help in making our early observations of GRB\,971214
possible.  We would also like to thank Bruce Margon, Christopher
Stubbs, and Eli Waxman for useful and informative discussions.  These
observations were made on very short notice and involved last-minute
technical changes which would not have been possible without the
enthusiastic cooperation of the APO site staff. We thank Daniel
Reichart for pointing out an error in the UT times of the APO $R$-band
observations in a pre-print of this paper, which when corrected, yield
a slower rate of decline. AD acknowledges the generous support of the
David and Lucille Packard Foundation.

\clearpage
 
\begin{deluxetable}{cccc}
\small
 
\tablecaption{Calibration stars used in the field of GRB\,971214.
  The uncertainties reflect photon statistics as well as uncertainty
  in the determination of the zero-point.\label{tbl-1}}
 
\tablewidth{0pt}
\tablehead{
\colhead{Star} & \colhead{$\alpha$(J2000)}  & \colhead{$\delta$(J2000)
}   & 
\colhead{$R$}} 
\startdata
1 & 11 56 25.8 & 65 11 36.3 & $19.96 \pm 0.03$ \nl
2 & 11 56 34.2 & 65 11 44.4 & $20.66 \pm 0.03$ \nl
3 & 11 56 29.8 & 65 12 16.3 & $16.38 \pm 0.03$ \nl
4 & 11 56 20.1 & 65 12 34.6 & $20.38 \pm 0.03$ \nl
5 & 11 56 13.4 & 65 11 15.6 & $20.50 \pm 0.03$ \nl
\enddata
\end{deluxetable}
 
\begin{deluxetable}{cccccl}
  \small 

  \tablecaption{Summary of APO observations of the GRB\,971214 optical
    afterglow.\label{tbl-2}}

\tablewidth{0pt}
\tablehead{
\colhead{Date (UT)} & \colhead{Days Since Burst} & \colhead{Exposures}   & \colhead{Seeing (arcsec)} & \colhead{$2 \sigma$ Detection Limit} & Magnitude $\pm  1\sigma$} 
\startdata
Dec 15.44 & 0.47 & $   730s$ & 1.3 & 21.8 & $J$ = $20.47^{+0.21}_{-0.19}$ \nl
Dec 15.51 & 0.54 & $  3300s$ & 1.4 & 24.6 & $R$ = $22.06 \pm 0.06 $       \nl
Dec 16.45 & 1.48 & $  1150s$ & 1.7 & 21.6 & $J$ = $21.46^{+0.34}_{-0.26}$ \nl
Dec 16.53 & 1.56 & $  1800s$ & 1.2 & 24.4 & $R$ = $23.36 \pm 0.13 $       \nl
Dec 17.50 & 2.53 & $  3000s$ & 1.2 & 24.8 & $R$ = $24.08 \pm 0.16 $       \nl
Dec 18.50 & 3.53 & $  3600s$ & 1.1 & 24.8 & $R$ = $24.61 \pm 0.22 $       \nl
\enddata
\end{deluxetable}


\end{document}